\documentclass[preprint,12pt,revtex4]{emulateapj}
\usepackage{xspace}
\usepackage{natbib}
\usepackage{verbatim}
\usepackage{graphicx,epstopdf}
\usepackage{longtable}
\usepackage{fancyref}
\usepackage{hyperref}
\usepackage{breakurl}
\usepackage{amsmath}
\usepackage{lscape}

\newcommand{\kep}{{\it Kepler}}
\newcommand{\ktwo}{{\it K2}\xspace}

\newcommand{\tess}{{\it TESS}\xspace}

\newcommand{\nhi}{\ensuremath{30}\xspace}
\newcommand{\nmed}{\ensuremath{48}\xspace}
\newcommand{\neb}{\ensuremath{164}\xspace}
\newcommand{\nvar}{\ensuremath{231}\xspace}
\newcommand{\ntce}{\ensuremath{1097}\xspace}
\newcommand{\nstars}{\ensuremath{20647}\xspace}

\providecommand{\adsurl}[1]{\href{#1}{ADS}}

\newcommand{\Teff}{$T_\mathrm{eff}$\xspace}
\newcommand{\logg}{$\log g$\xspace}

\newcommand{\ms}{\ensuremath{\rm m\,s^{-1}}}
\newcommand{\Kepler}{\textit{Kepler}\xspace} 
\newcommand{\TERRA}{\texttt{TERRA}\xspace}
\newcommand\nsf{$\dagger$}
\newcommand\hub{$\star$}
\newcommand\texaco{$\ddagger$}

\slugcomment{}

\shorttitle{K2 C16 Planet Candidates}
\shortauthors{Yu et al.}

\begin{document}


\title{Planetary Candidates from \ktwo Campaign 16}


\author{Liang Yu\altaffilmark{1},
Ian J.\ M.\ Crossfield\altaffilmark{1},
Joshua E.\ Schlieder\altaffilmark{2},
Molly R.\ Kosiarek\altaffilmark{3,\nsf},
Adina D.\ Feinstein\altaffilmark{4},
John H.\ Livingston\altaffilmark{5},
Andrew W.\ Howard\altaffilmark{6},
Bj{\"o}rn Benneke\altaffilmark{7},
Erik A.\ Petigura\altaffilmark{6,\hub},
Makennah Bristow\altaffilmark{8},
Jessie L.\ Christiansen\altaffilmark{9},
David R.\ Ciardi\altaffilmark{9},
Justin R.\ Crepp\altaffilmark{10},
Courtney D.\ Dressing\altaffilmark{11},
Benjamin J.\ Fulton\altaffilmark{7,\texaco},
Erica J.\ Gonzales\altaffilmark{3,\nsf},
Kevin K.\ Hardegree-Ullman\altaffilmark{12},
Thomas Henning\altaffilmark{13},
Howard Isaacson\altaffilmark{11},
S{\'e}bastien L{\'e}pine\altaffilmark{14},
Arturo O.\ Martinez\altaffilmark{14},
Farisa Y.\ Morales\altaffilmark{15,16},
Evan Sinukoff\altaffilmark{6,17}
}

\altaffiltext{1}{Department of Physics, and Kavli Institute for Astrophysics and Space Research, Massachusetts Institute of Technology, Cambridge, MA 02139, USA}
\altaffiltext{2}{NASA Goddard Space Flight Center, 8800 Greenbelt Road, Greenbelt, MD 20771, USA}
\altaffiltext{3}{Department of Astronomy and Astrophysics, University of California, Santa Cruz, CA 95064, USA}
\altaffiltext{4}{Department of Physics and Astronomy, Tufts University, Medford, MA 02155, USA}
\altaffiltext{5}{Department of Astronomy, Graduate School of Science, The University of Tokyo, Hongo 7-3-1, Bunkyo-ku, Tokyo, 113-0033, Japan}
\altaffiltext{6}{Cahill Center for Astrophysics, California Institute of Technology, Pasadena, CA 91125, USA}
\altaffiltext{7}{Division of Geological and Planetary Sciences, California Institute of Technology, Pasadena, CA 91125, USA}
\altaffiltext{8}{Department of Physics, University of North Carolina at Asheville, Asheville, NC 28804, USA}
\altaffiltext{9}{Caltech/IPAC-NASA Exoplanet Science Institute, 770 S. Wilson Ave, Pasadena, CA 91106, USA}
\altaffiltext{10}{Department of Physics, University of Notre Dame, Notre Dame, IN 46556, USA}
\altaffiltext{11}{Department of Astronomy, University of California, Berkeley, CA 94720, USA}
\altaffiltext{12}{The University of Toledo, 2801 West Bancroft Street, Mailstop 111, Toledo, OH 43606, USA}
\altaffiltext{13}{Max Planck Institute for Astronomy, K{\"o}̈nigstuhl 17, D-69117 Heidelberg, Germany}
\altaffiltext{14}{Department of Physics \& Astronomy, Georgia State University, 25 Park Place NE \#605, Atlanta, GA 30303, USA}
\altaffiltext{15}{Jet Propulsion Laboratory, California Institute of Technology, 4800 Oak Grove Drive, Pasadena, CA 91109, USA}
\altaffiltext{16}{Department of Physical Sciences, Moorpark College,7075 Campus Road, Moorpark, CA 93021, USA}
\altaffiltext{17}{Institute for Astronomy, University of Hawai'i, Honolulu, HI 96822, USA}

\altaffiltext{\nsf}{NSF Graduate Research Fellow}
\altaffiltext{\hub}{NASA Hubble Fellow}
\altaffiltext{\texaco}{Texaco Fellow}


\begin{abstract}
Given that Campaign 16 of the \ktwo mission is one of just two \ktwo campaigns observed so far in ``forward-facing'' mode, which enables immediate follow-up observations from the ground, we present a catalog of interesting targets identified through photometry alone.
Our catalog includes \nhi\ high-quality planet candidates (showing no signs of being non-planetary in nature), \nmed\ more ambiguous events that may be either planets or false positives, \neb\ eclipsing binaries, and \nvar\ other regularly periodic variable sources. 
We have released light curves for all targets in C16, and have also released system parameters and transit vetting plots for all interesting candidates identified in this paper.
Of particular interest is a candidate planet orbiting the bright F dwarf HD~73344 ($V=6.9$, $K=5.6$) with an orbital period of 15 days. If confirmed, this object would correspond to a $2.56 \pm 0.18 \ R_\oplus$ planet and would likely be a favorable target for radial velocity characterization.  
This paper is intended as a rapid release of planet candidates, eclipsing binaries and other interesting periodic variables to maximize the scientific yield of this campaign, and as a test run for the upcoming \tess mission, whose frequent data releases call for similarly rapid candidate identification and efficient follow-up.

\end{abstract}


\section{Introduction}

By any measure, NASA's {\em K2} mission \citep{howell:2014} has been a success.  Out of the ashes of an ailing spacecraft has risen a tremendously productive scientific mission.
Sometime this year, {\em K2} will likely run out of the propellant needed to   maintain its stable pointing and  deliver precise time-series photometry.  2018 is perhaps an appropriate year for this event,  since it marks the 40~year anniversary of the first American summit of K2 --- the ``Savage Mountain''.
Hundreds of planets and other astrophysical phenomena have been studied with {\em K2}, far fewer than the thousands discovered by the original {\em Kepler} mission \citep{thompson:2018} --- just as thousands of climbers have summited Mount Everest even though only hundreds have ever reached the top of K2.  Nonetheless, even after the mission ends, an enduring kinship will remain between those who have been fortunate enough to use {\em K2} in their research efforts.

In that same communal spirit, we provide  a rapid, public release of light curves, planet candidates, and other interesting periodic variables from {\em K2}'s Campaign 16 (C16) in this paper. Unlike most fields observed by {\em K2}, C16 was observed in ``forward-facing" mode, meaning that the field was observable throughout the night as soon as the campaign ended.  We have conducted a quick-look analysis of uncalibrated C16 cadence data and are releasing these data products in order to maximize the scientific yield of this campaign. We hope that this will also provide a test for the imminent \tess mission, whose frequent data releases will also benefit from rapid candidate identification and follow-up.

This paper is organized as follows:
in Sec.~\ref{k2data}, we describe how we compute time-series photometry and search for transit-like signals; Sec.~\ref{sec:vetting} then discusses our approach for discriminating between various astrophysical signals and measurement noise;  finally, in Sec.~\ref{sec:discussion} we conclude by discussing  several particularly interesting systems and reviewing the overall C16 candidate sample.

\section{\ktwo Targets and Photometry}\label{k2data}


\subsection{Target Selection and C16 Data Characteristics}\label{sec:data}
\ktwo target selection is entirely community driven, with all targets selected from Guest Observer (GO) proposals. Our team has proposed large samples of F, G, K, and M dwarfs for every \ktwo Campaign up to Campaign 17, but in the analysis that follows we use data from all \ktwo GO proposals to maximize the science yield from this campaign.

During C16, \ktwo observed \nstars\ stars in a field centered at RA = 08:54:50, Dec = +18:31:31, for a period of 80 days between 2017 Dec 07 and 2018 Feb 25. This is  only the second campaign in which the spacecraft was pointed along the forward-facing direction of  its velocity vector (the other, C9, was dedicated mostly to microlensing and was in a dense field unsuited for standard transit searches).  Forward-facing observations enable simultaneous observations from the ground and with {\em K2}, and also allow the field to be accessed from ground-based observatories as soon as compelling targets can be identified. C16 also overlaps with C5 except for a 40 px-wide strip that is not on silicon in C16. We find that 6167 targets observed in C16 were also observed in C5.

\subsection{Time-Series Photometry}\label{sec:phot}
Raw cadence pixel data for C16 became available on the Mikulski Archive for Space Telescopes (MAST)\footnote{https://archive.stsci.edu/k2/} on 2018 Feb 28. We first convert the raw cadence data into target pixel files with \texttt{kadenza}\footnote{https://github.com/KeplerGO/kadenza} \citep{kadenza}, following the approach described in \citet{christiansen18}. 

From then on we process the data using a photometric pipeline that has been described in detail in past works by members of our team \citep[e.g.][]{crossfield15, petigura15, petigura:2018}. In brief, we follow an approach similar to that originally outlined by \citet{vj14}. We compute the raw photometry by summing the flux within a soft-edged, stationary, circular aperture centered on each target star. During \ktwo operations, solar radiation pressure causes the telescope to roll around its boresight. Consequently, stars trace out small arcs of up to several pixels every $\sim6$ hr. Interpixel sensitivity variations and aperture losses can then lead to significant changes in the brightness of stars that dominate \ktwo photometry.

\begin{figure*}[htb]
\begin{center}
\includegraphics[width=0.8\textwidth]{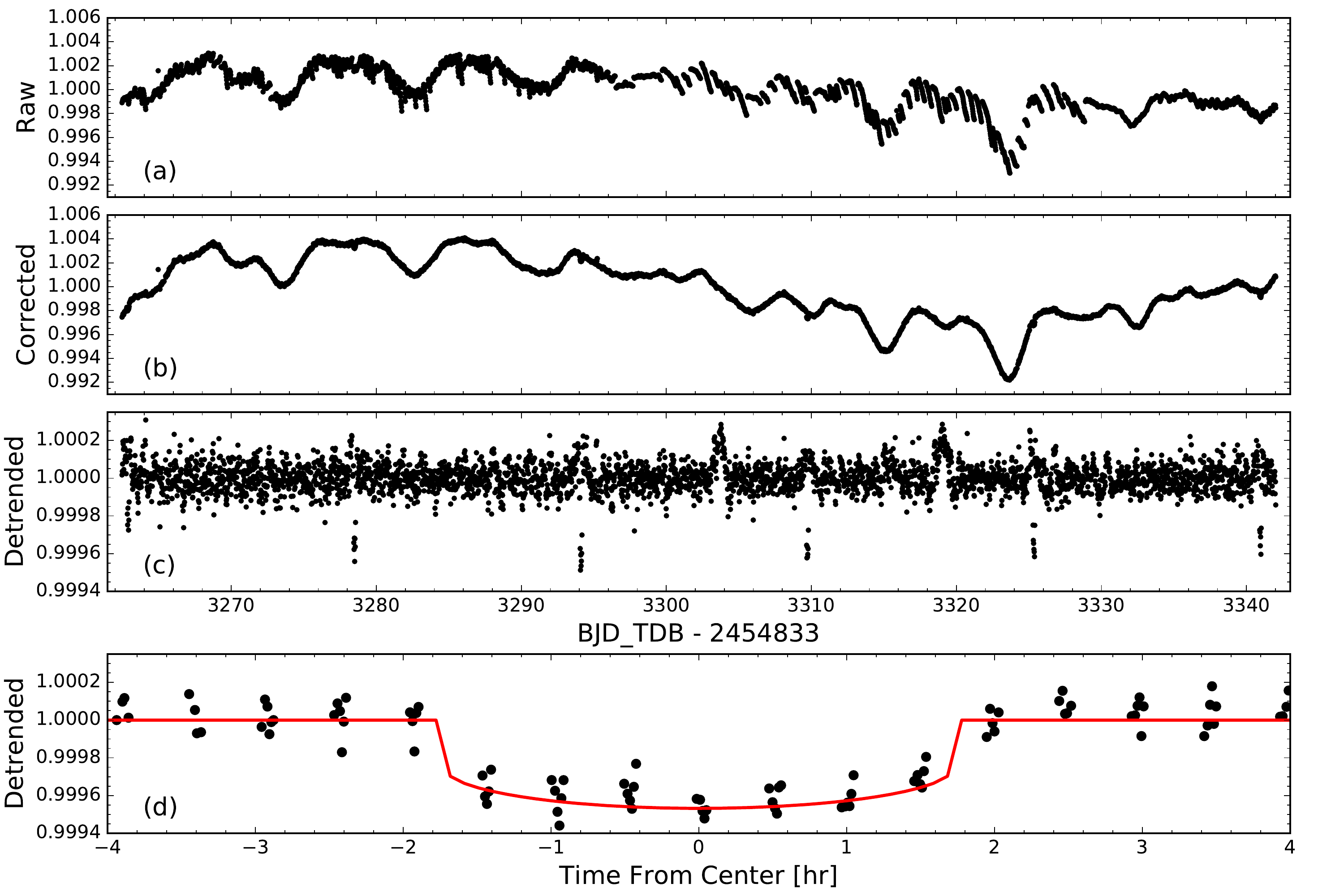}
\caption{\label{fig:lc} {\em K2} photometry of HD~73344 ($V=6.9$) and its planet candidate, EPIC 212178066.01. From top to bottom: raw aperture photometry; after removal of telescope systematics, revealing a likely $8.5\pm0.5$~d rotation period; after detrending, clearly revealing  candidate transits; and after phase-folding and overplotting  a model transit profile (red). The bumps in panel (c) do not occur on the same period as the candidate transit and may be artifacts of the detrending process.}
\end{center}
\end{figure*}

To correct for these motion-dependent systematics, we solve for the roll angle between each frame and an arbitrary reference frame using roughly 100 stars of Kepler magnitude $Kp \sim 12$ mag on an arbitrary output channel (we typically use channel 4). Then we use the publicly available \texttt{k2phot} photometry code\footnote{https://github.com/petigura/k2phot} to model the time- and roll-dependent brightness variations using a Gaussian process with a squared-exponential kernel. The models can be individually  applied to  the raw photometry to produce photometry corrected for motion-dependent systematics or fully detrended photometry. Fig.~\ref{fig:lc} shows an example of raw \ktwo photometry for a relatively well-behaved star, along with the same light curve after correction for systematics and subsequent detrending. Some light curves with relatively deep transits, as in this example, show small increases in flux immediately before and after the transits. These are artifacts from the detrending process. The transits are effectively outliers on short timescales that may bias the Gaussian process model, leading to overfitting.

We repeat this photometry process for apertures with radii ranging from 1 to 7 pixels, and also fit a custom, automatically-generated aperture that selects pixels based on how much flux they receive relative to the background. This aperture has an irregular shape and captures most of the flux from each target. For each target we adopt the aperture  that minimizes the residual noise on 3~hr timescales. Specifically, we use the median absolute deviation (MAD) of the 3~hr Single Event Statistic (SES) as our noise metric. We define the SES as the depth of a box-shaped dimming relative to the local photometric level. This method of aperture selection favors smaller apertures, which incur less background noise, for fainter stars and larger apertures for brighter targets. For strongly saturated stars the custom aperture is typically chosen, since  in these cases circular apertures  miss substantial flux.


\subsection{Transit Search}\label{sec:tps}
We search our calibrated photometry for transit signals using the publicly available \texttt{TERRA} algorithm\footnote{https://github.com/petigura/terra} \citep{petigura13a,petigura13b}. \texttt{TERRA} flags targets with putative transits as threshold-crossing events (TCEs), which we later examine visually (see Sec.~\ref{sec:vetting}). Once a TCE is detected,  \texttt{TERRA} automatically runs again to search for additional signals in the same system  \citep[see][]{petigura:2018} until no more TCEs are found or until the number of candidates exceeds 5. 

Many spurious detections at lower S/N are caused by residual outliers in the photometry. In order to reduce the number of spurious detections, we require that TCEs have orbital periods longer than 0.5~d, and that they also show at least three transits. This last criterion rules out any planets with periods longer than half the campaign baseline, or $\sim 40$ days. Thus many longer-period planets likely remain to be found in this data set. Furthermore, we adopt a threshold of S/N$\ge12$ to yield a good balance between sensitivity to shallow transits and the number of spurious detections. In previous catalog papers produced using the fully processed target pixel files released later by the {\em K2} project office, we typically vetted candidates down to a lower S/N threshold of 10. We find that spurious detections are more frequent in light curves derived from uncalibrated cadence data than when using fully calibrated pixel files. 

\begin{figure}[htb]
\begin{center}
\includegraphics[width=0.5\textwidth]{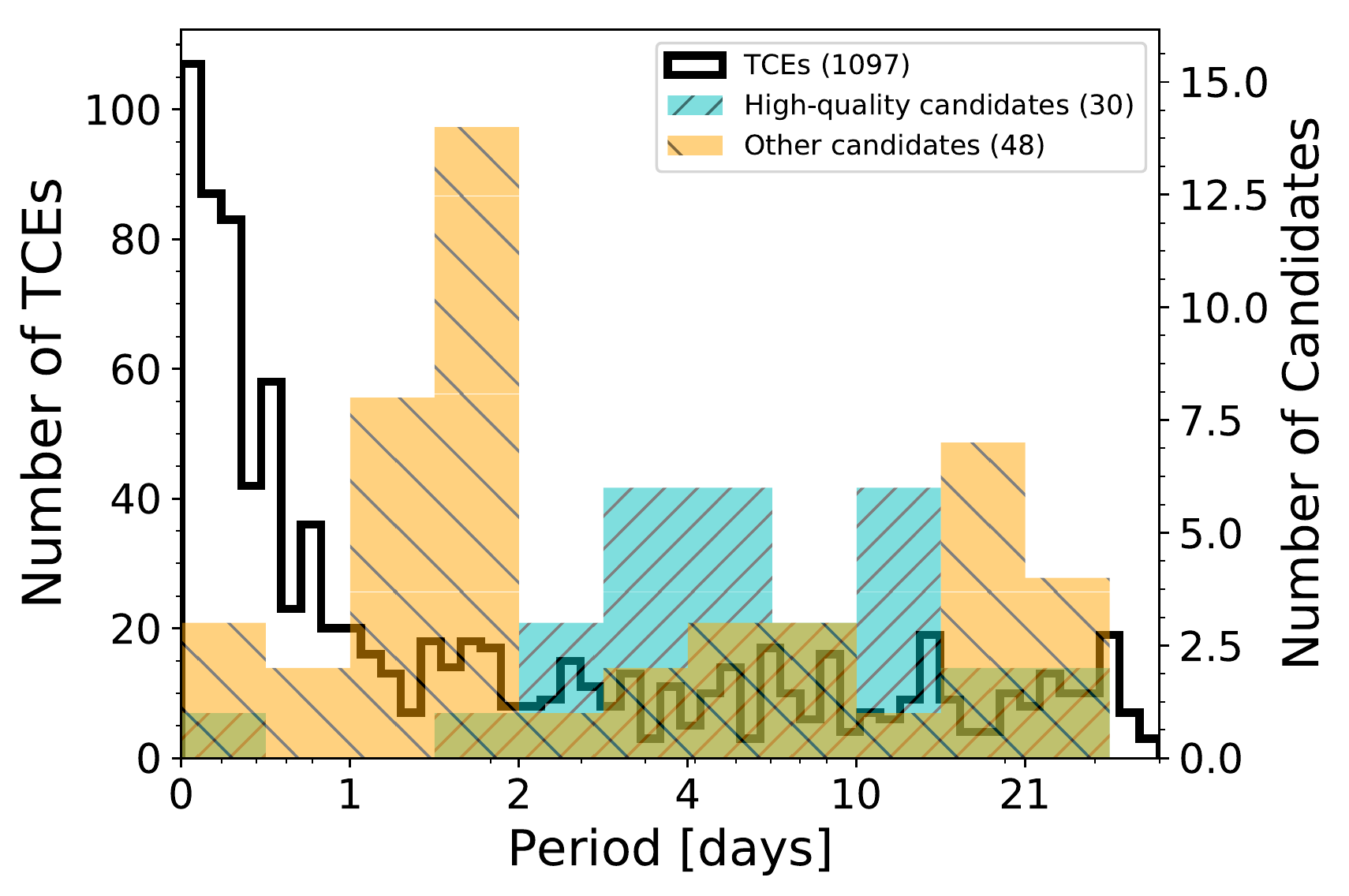}
\caption{\label{fig:per_dist} Orbital periods of transit-like
  signals identified in our analysis. The unfilled, narrow-binned histogram (axis at
  left) shows the Threshold-Crossing Events (TCEs) identified by
  \TERRA in our initial transit search (see Sec.~\ref{sec:tps}).  The
  coarser histograms (axis at right) indicate the distributions of
  \nhi high-quality candidates (blue-green) and \nmed remaining, plausibly planetary candidates
  (orange). }
\end{center}
\end{figure}

In total, \texttt{TERRA} produced a list of \ntce\ TCEs in C16 with nominal S/N$\ge$12. The distribution of their orbital periods is shown in Fig.~\ref{fig:per_dist}.

\section{Triage and Vetting}\label{sec:vetting}
The majority of TCEs identified by \texttt{TERRA} are not caused by genuine transiting planets, but instead by residual instrumental artifacts, eclipsing binary stars, or other periodic stellar variability (e.g. pulsations and spot modulations).  We  manually vet our entire list of \ntce\ TCEs to differentiate between these various signals. This process results in a list of robust planet candidates for further follow-up and validation, as well as a list of eclipsing binaries and other periodically variable sources.

We promote TCEs showing no obvious warning signs to the status of ``planet candidate'' in the spirit of ``Kepler Objects of Interest'' (KOIs), i.e.\  events that are almost certainly astrophysical in nature and not obviously  false positive scenarios such as eclipsing binaries or variable stars. Details of the vetting process are described in \citet{crossfield16} and \citet{petigura:2018}. \texttt{TERRA} produces a set of diagnostics for every TCE, which we use to classify the event as a candidate planet, eclipsing binary, periodic variable, or noise. The diagnostics include a summary of basic fit parameters and a suite of diagnostic plots to visualize the nature of the TCE.  These plots include the \texttt{TERRA} periodogram, a normalized phase-folded light curve with the best-fit \citet{mandelagol} model, the light curve phased to $180^{\circ}$ to look for eclipses or misidentified periods, the most probable secondary eclipse identified at any phase, and an autocorrelation function. In the era of \tess, cross-matching to ground-based surveys will be another excellent way to discover false positives \citep[e.g.][]{oelkers18}. 

\begin{figure*}[htb]
\begin{center}
\includegraphics[width=\textwidth]{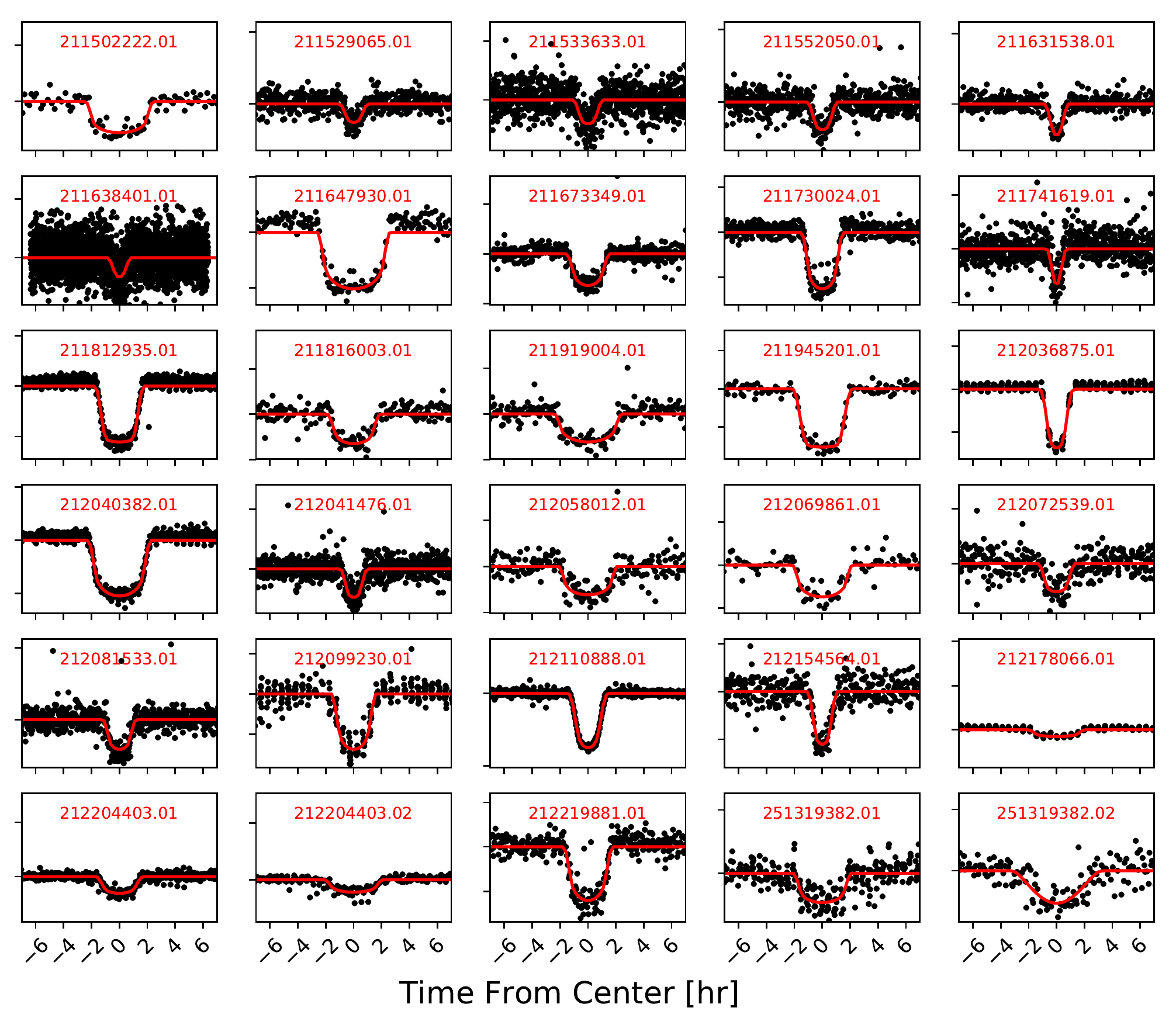}
\caption{\label{fig:hi_cand} Phase-folded light curves of our \nhi\ high-quality planet candidates, and their best-fit \citet{mandelagol} transit models. To avoid clutter, we did not label the y-axis. Their system parameters are listed in Table~\ref{tab:good}.   }
\end{center}
\end{figure*}

Table~\ref{tab:good} lists the \nhi\ highest-quality planet candidates whose light curves (shown in Fig.~\ref{fig:hi_cand}) show no obvious signs of being non-planetary in nature; our experience with four years of {\em K2} data leads us to believe that most of these are indeed real planets, ready to  be confirmed (e.g., via mass measurements) or statistically validated. 

\begin{figure*}[htb]
\begin{center}
\includegraphics[width=\textwidth]{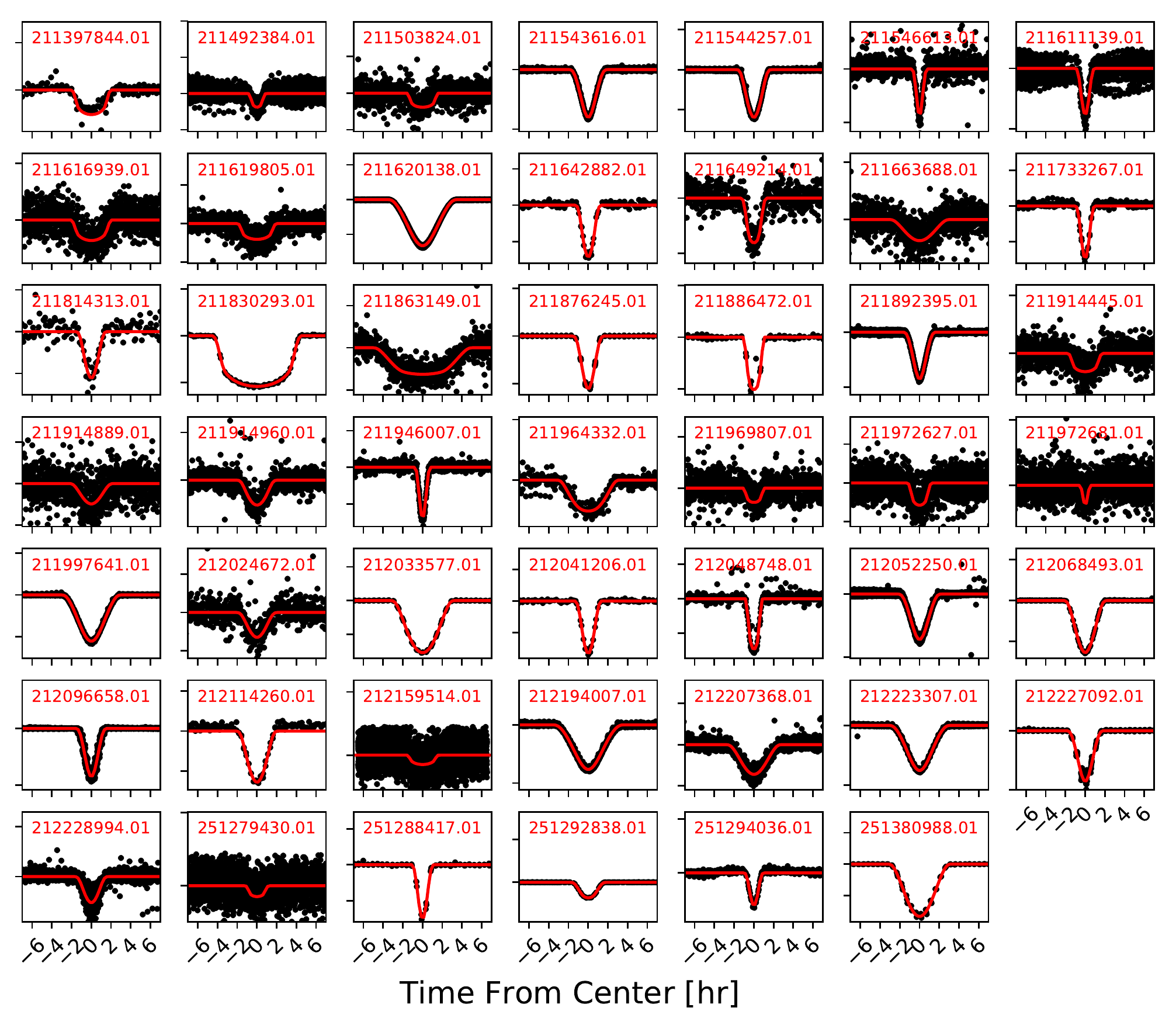}
\caption{\label{fig:med_cand} Phase-folded light curves of our \nmed\ lower-quality planet candidates, and their best-fit \citet{mandelagol} transit models. Typical transit depths for these candidates range from 300 ppm to 700000 ppm. Their system parameters are listed in Table~\ref{tab:med}.   }
\end{center}
\end{figure*}

Table~\ref{tab:med} lists  \nmed\  candidates that could also be transiting planets but include some ambiguous warning signs such as a V-shaped transit (frequently caused by eclipsing binaries).  Some candidates in this list may be real planets, but many are likely non-planetary.  Following the examples of the KOIs and of \citet{vanderburg16}, we do not classify candidates with very deep transits as false positives  even though transit depths  $\gtrsim5\%$  very likely indicate eclipsing binaries. Candidates with radii larger than 1.5 $R_J$ were also included in this category, since giant planet candidates from \kep\ have a  false positive rate as high as 50\% \citep{santerne16}. We plot the light curves of these candidates in Fig.~\ref{fig:med_cand}.


Finally, we identify a larger sample of periodic astrophysical signals that are almost certainly not caused by planets.  Table~\ref{tab:eb} lists  \neb\  targets that clearly show both transits and secondary eclipses, while Table~\ref{tab:var} lists the \nvar\  other periodic, astrophysical signals such as pulsations, coherent stellar rotations, and objects identified as galaxies or quasars in the Ecliptic Plane Input Catalog \citep[EPIC;][]{huber16} or GO proposals.  There is likely overlap between these last two tables, e.g.\ for  short-period contact/near-contact binaries whose light curves may have been classified as periodic variables.


After constructing the samples of astrophysical TCEs described above, we also perform ephemeris matching following the approach of \cite{coughlin14}.  By adopting their  recommended thresholds for  periods and times-of-transit, we identify a number of transit-like signals with matching ephemerides.    We do not discard any of these systems, but indicate them in our target tables. This matching exercise also led us to demote 2 systems that we had originally classified as high-quality candidates (211914445.01 and 211964332.01) down into a lower tier.

To provide the community access to these candidates as rapidly as possible, we have chosen to forego a full MCMC analysis on each candidate's light curve. Instead, we run a Levenberg-Marquardt minimization on each planet candidate to fit a \citet{mandelagol} transit model. The stellar limb darkening parameters are fixed to values derived using the \texttt{PyLDTk} package\footnote{https://github.com/hpparvi/ldtk/tree/v1.0} \citep{ldtk} and stellar parameters derived in Section~\ref{sec:stellar}. We find that this fit gives us a more reliable estimate of the transit ephemerides than \texttt{TERRA}. For periodic variables and systems with secondary eclipses, we merely report the parameters found by \texttt{TERRA}.
In some cases \texttt{TERRA} obviously identified a multiple of the true period, and we include a note to that effect where appropriate.

\section{Discussion}\label{sec:discussion}
\subsection{Host Star Parameters} \label{sec:stellar}
Unlike the original \Kepler mission, \ktwo does not have a homogeneous  catalog of stellar parameters. Fortunately, we still have the benefit of the comprehensive classification catalog of \ktwo targets produced by \citet{huber16}, who used mainly a combination of colors and galactic population synthesis models to derive stellar parameters such as effective temperatures (\Teff), surface gravities (\logg), metallicities ([Fe/H]), radii, masses, densities, distances and extinctions for \ktwo stars. The typical precision of these classifications is $\approx2\%-3\%$ in \Teff \citep{huber16}.  However, the \citet{huber16} analysis misclassifies 55-70\% of subgiants as dwarfs, and relies on Padova stellar models \citep{marigo08}, which systematically underestimate the stellar radii of M dwarfs by up to 20\%. Many C16 targets, including all of our planet candidate hosts, also have parallaxes from Gaia DR2 \citep{gaia:2016, gaia:2018}. We used the parallaxes and the \texttt{isochrones} package\footnote{https://github.com/timothydmorton/isochrones} \citep{isochrones} in conjunction with the broadband photometry ($BVJHKgri$) from the EPIC to infer the \Teff, stellar radii, \logg, [Fe/H] and masses of all planet candidate hosts. In Tables~\ref{tab:good} and~\ref{tab:med}, we list the median stellar parameters and their $1\sigma$ uncertainties from \texttt{isochrones} for all of our candidates. For the vast majority of the candidates, the best-fit \Teff is consistent with that from \citet{huber16} at the $2\sigma$ level. But we note that the reported uncertainties are only statistical uncertainties and do not account for any systematic uncertainties in the underlying stellar models, and may therefore be underestimated, especially for cooler stars.

\begin{figure}[ht]
\begin{center}
\includegraphics[width=0.5\textwidth]{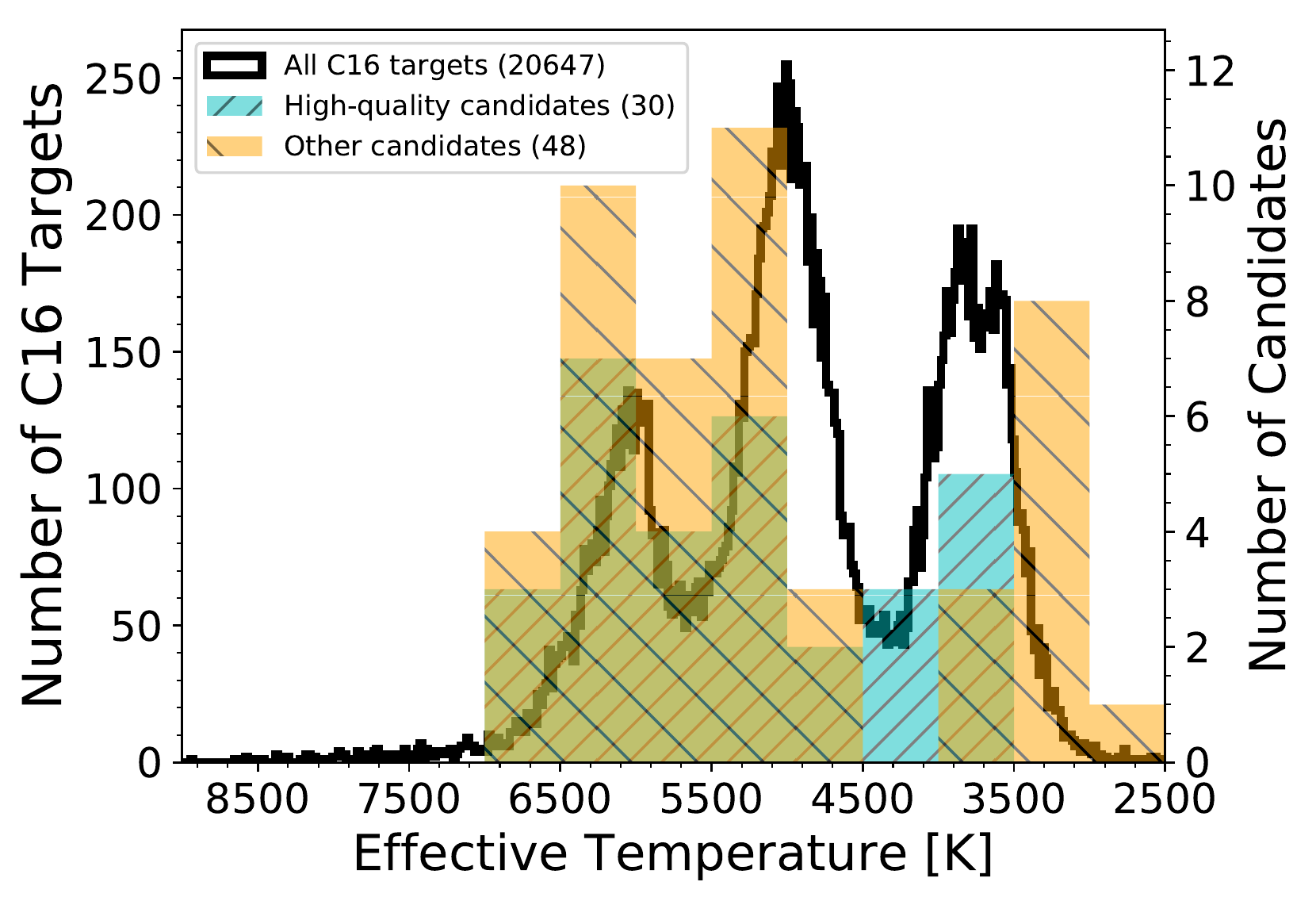}
\caption{\label{fig:teff_dist}  Distribution of EPIC stellar $T_\mathrm{eff}$ for the entire C16 target sample (empty, fine-grained histogram) and for our planet candidate sample (filled, coarser histograms). }
\end{center}
\end{figure}

Fig.~\ref{fig:teff_dist} shows the \citet{huber16} \Teff (where available) for the entire C16 sample, along with the \texttt{isochrones}-derived \Teff distribution among our planet candidate samples. The full campaign shows three distinct populations of targets observed by {\em K2}, with peaks around 3500~K, 5000~K, and 6100~K.  The number of candidates is of course much lower, but the distribution of \Teff for these systems appears to roughly track that of the underlying target distribution even though we do not expect it to, given the change in planet detectability as a function of stellar magnitude, radius, and noise.

\subsection{Characteristics of the Planet Candidate Sample}
 \label{sec:sample}
 The period distribution of our planet candidates, along with that of the TCEs, is shown in Fig.~\ref{fig:per_dist}.  Whereas the TCE distribution peaks for $P<1$~d, the number of high-quality candidates increases towards longer periods as expected for real planets \citep[e.g.,][]{morton16,fulton:2017}. A larger fraction of lower-quality candidates have $P<2$~d; based on the occurrence rates of short-period planets,  we expect that many of these shortest-period candidates are not planets.

\begin{figure}[htb]
\begin{center}
\includegraphics[width=0.5\textwidth]{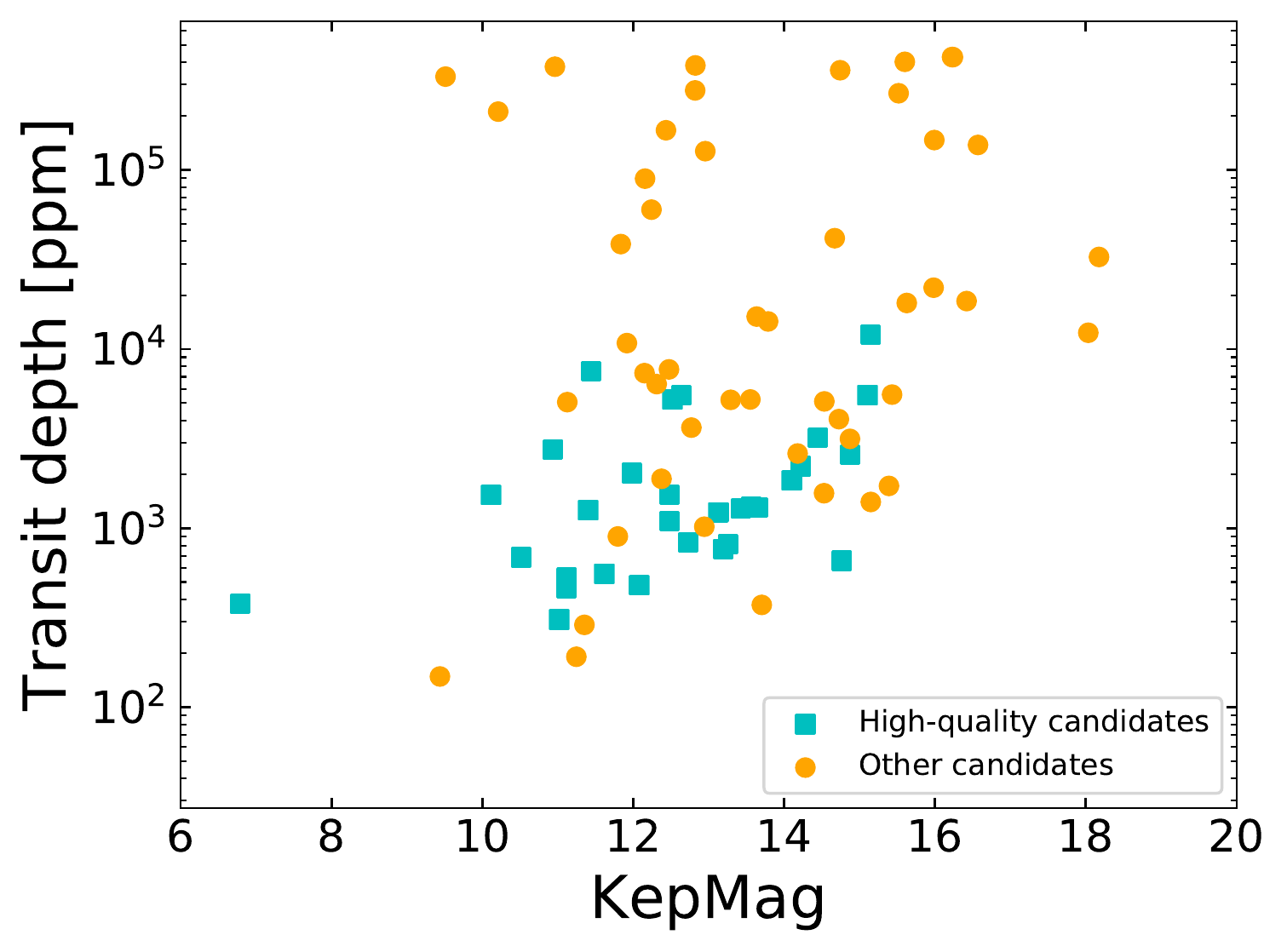}
\caption{\label{fig:kepmag-trandep}  Transit depth and
  stellar magnitude for our high-quality candidates (light blue squares) and lower-quality candidates (orange circles).}
\end{center}
\end{figure}

Fig.~\ref{fig:kepmag-trandep} shows the brightness in the \kep\ bandpass ($Kp$) and transit depths for our candidates. The highest-quality candidates typically orbit stars with $Kp=10-15$~mag and have transit depths $\gtrsim$~100~ppm, as is typical for {\em K2} planet catalogs \citep[e.g.,][]{mayo:2018}.  One candidate has  $Kp=6.8$~mag and is a clear outlier; this would be the brightest host star, by far, for any transiting planet discovered by {\em K2}.  We discuss this candidate, HD~73344, in more detail in Sec.~\ref{sec:hd} below. 

\begin{figure}[htb]
\begin{center}
\includegraphics[width=0.5\textwidth]{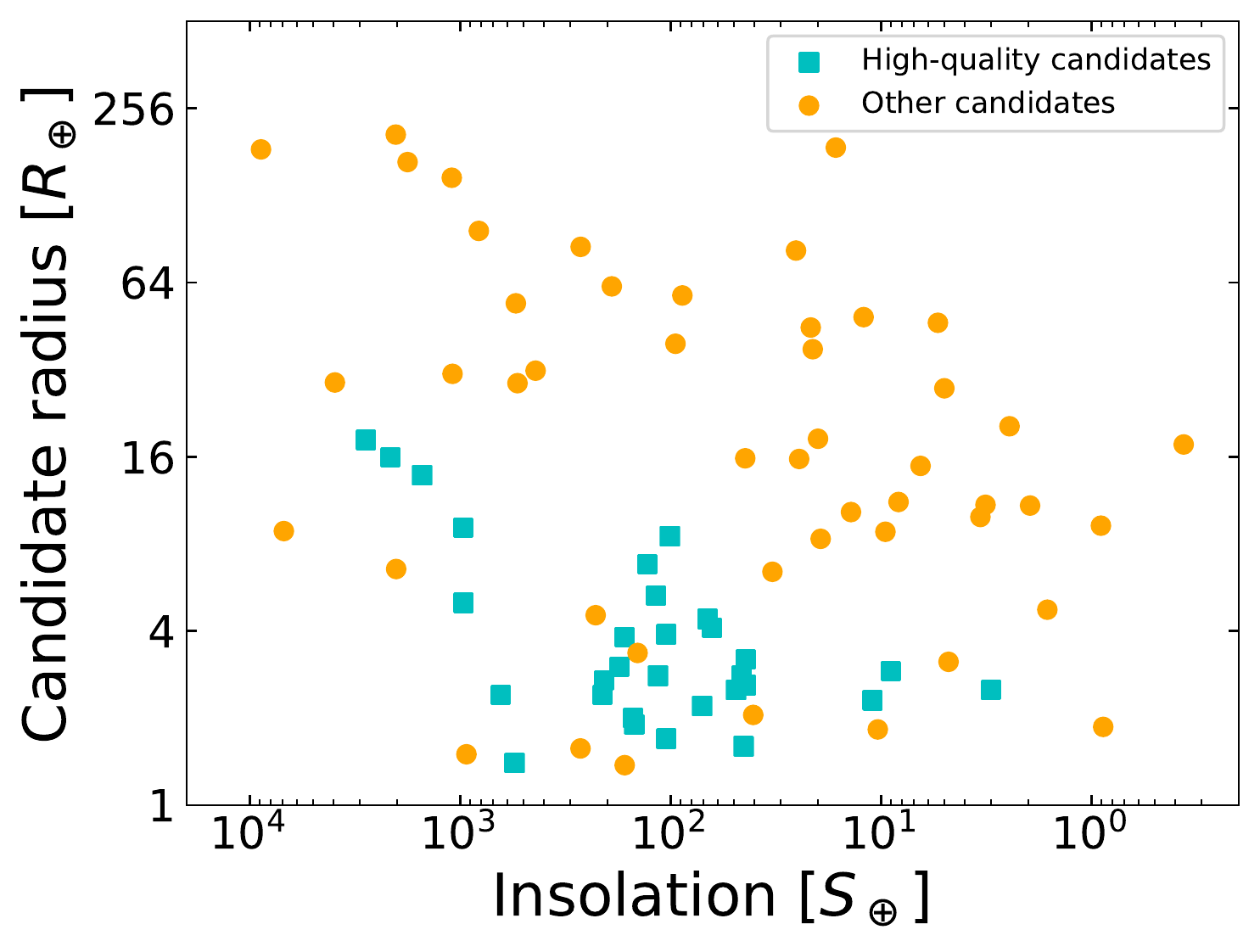}
\caption{\label{fig:insol-rad} Approximate radii and incident
  insolation for our high-quality candidates (light blue squares) and lower-quality candidates (orange circles).   }
\end{center}
\end{figure}

Adopting the stellar parameters derived in Section~\ref{sec:stellar}, Fig.~\ref{fig:insol-rad} plots the planet radii and incident irradiation of all our candidates. 

We detect two possible multi-planet systems. Two high-quality candidates orbit EPIC~212204403, with periods of~4.7 and~12.6~d and sizes of approximately~3.3 and~2.6~$R_\oplus$, respectively.  Another two high quality candidates are detected around EPIC~251319382, with periods of~8.2 and~14.9 days, and radii of~2.0 and~4.4~$R_\oplus$. Based on past studies of multi-planet systems these candidates are likely to be real planets \citep{lissauer:2012,sinukoff16}. Validating them is beyond the scope of this work, but at $V=11-12$ mag, these systems could be interesting targets for radial velocity (RV) mass measurements of multi-planet systems.

Another interesting candidate is EPIC~212048748.01 from the lower-quality ``plausible planet candidate" list. This candidate transits with a 3155 ppm depth and a period of 5.75~d around a high proper motion, infrared bright ($K = 9.2$) star having optical-IR photometry consistent with an M3 spectral type. If confirmed, this $\sim2 R_{\oplus}$ candidate will be a priority target for upcoming IR sensitive precision RV instruments and transit spectroscopy with the {\it James Webb Space Telescope}.

Finally, a comparison with the NASA Exoplanet Archive shows that four of our candidates have already been validated using data from C5.  \cite{dressing17} validated two of our high-quality C16 candidates,  212069861.01 (K2-123b) and 212154564.01 (K2-124b); another candidate 212110888.01 is a previously known hot Jupiter K2-34b \citep{lillobox16, hirano16}; and our lower-priority candidate 211969807.01 was validated as K2-104b \citep{mann:2017}.  One more low-quality candidate, 211946007.01, was confirmed to be a transiting brown dwarf \citep{gillen17}. Our derived system parameters are in approximate agreement with those in the discovery papers. A combined analysis  of the C5 and C16 data (possible for many targets in C16) may prove fruitful for these systems.

\subsection{HD~73344}
\label{sec:hd}
One candidate of particular interest is HD~73344\footnote{This target was proposed by many K2 GO programs:  16009 (PI Charbonneau), 16010 (PI Lund), 16021 (PI Howard), 16028 (PI Cochran), 16063 (PI Redfield), 16068 (PI Jensen), and 16081 (PI Guzik).} (HIP~42403, EPIC~212178066), and we show the light curve in Fig~\ref{fig:lc}.  This bright F star ($V=6.9$~mag) is highly saturated in the \ktwo data, but a custom aperture encompassing the entire saturated PSF shows the clear transit-like signal highlighted in Fig.~\ref{fig:lc}.  Because the candidate is exceptionally bright, and thus amenable to future characterization, we investigated the signal more closely than others, as explained below.

The star has been characterized by many groups over the years \citep[e.g.,][]{valenti:2005,paletou:2015}.   It lies at a distance of $35.296\pm0.052$~pc \citep{gaia:2018} and its parameters are \Teff$=6120\pm50$~K, $R_*=1.15\pm0.04\, R_\odot$, $M_*=1.26\pm0.19\, M_\odot$ \citep{valenti:2005}, in good agreement with our derived values from Gaia DR2 and \texttt{isochrones}.  The star's projected rotational velocity is $v \sin i=6.3\pm0.5$~km~s$^{-1}$ \citep{valenti:2005}, and our light curve shows evidence of stellar rotation at a period (determined via  Lomb-Scargle periodogram)  of $8.5\pm0.5$~d. 
This period would be consistent with the rotation periods of other stars with similar colors, and is consistent with a stellar age of roughly 1~Gyr \citep{angus:2015}. Combining all these parameters indicates that the stellar rotation axis is inclined by $i=62^\circ\pm10^\circ$. Thus, if the candidate signal comes from an object orbiting HD~73344, the angular momentum of the star and the transiting object's orbit are likely misaligned.


Because the star is strongly saturated, we cannot apply a standard centroid analysis of the stellar position in- vs. out-of-transit. However, a transit analysis with MCMC \citep[identical to that described by][]{crossfield15}    implies a stellar density of $\rho_{*,circ}=2.2\pm1.2$~g~cm$^{-3}$ --- a loose constraint, but consistent with the spectroscopically-inferred stellar density of $1.2\pm0.2$~g~cm$^{-3}$ and much higher than the low stellar densities that might be expected from an eclipsed giant star.  The results of our transit analysis, which includes dilution as a free parameter, are also consistent with no dilution. 

The resulting parameters from our transit analysis of HD~73344 are listed in Table~\ref{tab:planet}. If the transits are occurring around the target and not around a background star in the photometric aperture, the stellar radius and transit depth imply a candidate radius of roughly 2.6~$R_\oplus$. This size would imply a corresponding candidate mass of $10\pm3\,M_\oplus$ \citep{wolfgang:2016} and an RV amplitude of $\sim$2~m~s$^{-1}$.  The star was observed 24 times over eleven years as part of the Lick radial velocity survey \citep{fischer:2014}, but these data have an RMS of 32~m~s$^{-1}$ (despite internal uncertainties of roughly 6~m~s$^{-1}$) and  show no coherent RV signal at the candidate period or at our calculated stellar rotation period.  Nightly Keck/HIRES RVs over four consecutive nights in 1999 showed a stellar jitter  of   3.9~\ms\ \citep{isaacson10}. HD~73344 also exhibits moderate chromospheric activity \citep[$S_{HK}=0.22$, $R'_{HK}=-4.66$;][]{isaacson10}, but at this \Teff H\&K activity is not the main contribution to jitter. 
 It seems likely that precise RV measurements could confirm this planet candidate, despite the fact that it orbits an early-type star, which makes RV measurements more challenging than for later-type stars.

\begin{deluxetable}{l l l}[bt]
\tabletypesize{\scriptsize}
\tablecaption{Candidate Parameters for HD~73344 \label{tab:planet}}
\tablehead{
\colhead{Parameter} & \colhead{Units} & \colhead{Value} 
}
   $T_{0}$ & $\mathrm{BJD_{TDB}} - 2454833 $ & $3262.8931^{+0.0020}_{-0.0023}$ \\
       $P$ &          d & $15.61335^{+0.00085}_{-0.00078}$ \\
       $i$ &        deg & $89.15^{+0.61}_{-1.13}$ \\
 $R_\mathrm{circ}/R_*$ &         \% & $2.65^{+0.15}_{-0.10}$ \\
  $T_{14}$ &         hr & $3.46^{+0.20}_{-0.17}$ \\
  $T_{23}$ &         hr & $3.22^{+0.21}_{-0.18}$ \\
   $R_*/a$ &         -- & $0.0327^{+0.0118}_{-0.0042}$ \\
       $b$ &         -- & $0.46\pm0.32$ \\
$\rho_{*,\mathrm{circ}}$ & g~cm$^{-3}$ & $2.2^{+1.1}_{-1.3}$ \\
       $a$ &         AU & $0.1321^{+0.0063}_{-0.0070}$ \\
     $R_\mathrm{cand}$ &      $R_E$ & $2.56\pm0.18$ \\
 $S_\mathrm{inc}$ &      $S_E$ & $111^{+12}_{-11}$ 
\enddata
\end{deluxetable}

\subsection{Conclusions}
\label{sec:conclusions}
In a short timespan, we have converted cadence-level {\em K2} data into time-series photometry of \nstars\ targets, identified  \ntce\ periodic signals (of astrophysical or instrumental origin), and distilled these into \nhi\ high-quality planet candidates, \nmed\ lower-quality candidates, \neb\ eclipsing binaries, and \nvar\ other periodically-variable astrophysical sources.  Four of our candidates have already been validated as planets (see Sec.~\ref{sec:sample}), suggesting that our approach successfully identifies planet-like signals.  One particularly interesting new target is HD~73344, a $V=6.9$ F dwarf which may host a 2.6~$R_\oplus$ planet on a 15~d orbit (see Sec.~\ref{sec:hd}).  We have released parameters for all identified systems of interest, along with light curves and transit vetting plots\footnote{All available now at \url{https://exofop.ipac.caltech.edu/k2/}, or by request.}. We hope that rapid identification and public dissemination of interesting signals will maximize the scientific productivity of  {\em K2}. If {\em K2} continues operating through the end of C17 (another forward-facing campaign), it may prove useful to perform a similarly rapid analysis of those data.

This rapid-release model is also somewhat of an  analog for the upcoming {\em TESS} mission \citep{ricker:2014}. The release of planet catalogs has occurred only irregularly during the {\em K2} mission, but this paradigm will change once {\em TESS} operations begin in earnest.  Data from {\em TESS} will be released and processed  on a 27-day rhythm  for most of the  two-year mission duration. With the shorter observing windows, ephemeris decay is also a much larger problem for \tess and therefore the importance of securing planet candidates in the same season is even higher. If interesting objects could be rapidly gleaned from {\em TESS} data and circulated to the community, follow-up observations and analyses could begin a full season earlier and so the full impact of that mission could more quickly be achieved.

\acknowledgments
We thank the anonymous referee and Trevor David for providing helpful comments on the manuscript, and all those who selected the targets observed in C16.  I.J.M.C.\ acknowledges support from NASA through  K2GO grant 80NSSC18K0308 and from NSF through grant AST-1824644. He also gratefully acknowledges the hospitality of the organizers and participants of the ``Challenge to Super-Earths'' workshop at NAOJ, during which much of this work took place. This work made use of the gaia-kepler.fun crossmatch database created by Megan Bedell.
This paper includes data collected by the \kep\ mission. Funding for the \kep\ mission is provided by the NASA Science Mission directorate. Some of the data presented in this paper were obtained from the Mikulski Archive for Space Telescopes (MAST). STScI is operated by the Association of Universities for Research in Astronomy, Inc., under NASA contract NAS5--26555. Support for MAST for non--HST data is provided by the NASA Office of Space Science via grant NNX13AC07G and by other grants and contracts. This research has made use of the Exoplanet Follow-up Observing Program (ExoFOP), which is operated by the California Institute of Technology, under contract with the National Aeronautics
and Space Administration.

Facilities: \facility{Kepler}, \facility{K2}
\bibliographystyle{apj_hyperref}
\bibliography{main}

\clearpage
\begin{turnpage}
\begin{deluxetable}{lllllllllllllp{2cm}}
\tabletypesize{\scriptsize}\tablecaption{High-quality candidate parameters}
\tablehead{    Candidate &  $Kp$ &      $P$ &       $T_0$ & $T_{14}$ &  $(R_P/R_*)^2$ & $R_*$\tablenotemark{a} & $M_*$\tablenotemark{a} & \Teff\tablenotemark{a} & \logg\tablenotemark{a} & $R_P$ &  $S_\mathrm{inc}$ & In C5? & Comment\\
& & [d] & [BJD - 2454833] & [d] & [ppm] & [$R_\odot$] & [$M_\odot$] & [K] & [cgs] & [$R_\oplus$] & [$F_\oplus$] }\startdata

 211502222.01              & 11.617 & 22.995920 & 3280.30425 & 0.155 & 470 & $1.043 \pm 0.015$ & $1.054 \pm 0.060$ & $6010 \pm 170$ & $4.426 \pm 0.031$ &  2.5 & 49 &  &                  --- \\
 211529065.01              & 13.431 &  4.399871 & 3264.73662 & 0.0532 & 1076 & $0.772 \pm 0.006$ & $0.852 \pm 0.019$ & $4810 \pm 30$ & $4.594 \pm 0.010$ &  2.8 & 115 & yes &                  --- \\
 211533633.01              & 13.258 &  2.126953 & 3262.66212 & 0.059 & 707 & $0.573 \pm 0.018$ & $0.585 \pm 0.024$ & $4020 \pm 220$ & $4.696 \pm 0.015$ &  1.7 & 105 &  &                  --- \\
 211552050.01              & 13.194 &  3.544965 & 3264.95164 & 0.061 & 621 & $0.701 \pm 0.007$ & $0.718 \pm 0.037$ & $4860 \pm 120$ & $4.602 \pm 0.024$ &  1.9 & 149 &  &                  --- \\
 211631538.01              & 14.221 &  5.513327 & 3266.24778 & 0.0386 & 2119 & $0.770 \pm 0.014$ & $0.791 \pm 0.046$ & $5020 \pm 110$ & $4.562 \pm 0.030$ &  3.9 & 105 &  &                  --- \\
 211638401.01              & 14.763 &  0.530270 & 3262.64176 & 0.0443 & 568 & $0.558 \pm 0.006$ & $0.578 \pm 0.013$ & $3880 \pm 70$ & $4.706 \pm 0.011$ &  1.4 & 553 &  &                  --- \\ 
 211647930.01              & 11.982 & 14.759595 & 3264.39324 & 0.1712 & 1705 & $1.186 \pm 0.024$ & $1.068 \pm 0.079$ & $6070 \pm 190$ & $4.318 \pm 0.043$ &  5.3 & 118 &  &                  --- \\
  211673349.01              & 14.446 &  4.894289 & 3264.62726 & 0.0877 & 3043 & $0.686 \pm 0.008$ & $0.712 \pm 0.026$ & $4440 \pm 50$ & $4.620 \pm 0.018$ &  4.1 & 64 &  &                  --- \\
\enddata\label{tab:good}
\tablenotetext{a}{Note that all uncertainties reported in this table and Table~\ref{tab:med} are statistical uncertainties only and do not account for systematic uncertainties in the underlying stellar models.}
\tablenotetext{b}{Parameters not in EPIC; classified using \texttt{isochrones} as described in Sec.~\ref{sec:stellar}.}
\tablenotetext{c}{We identified hints of a third candidate in this system, with a period of $\sim 3.5$ d and S/N$\sim$7.}
\end{deluxetable}

\end{turnpage}
\clearpage

\begin{landscape}
\LongTables
\begin{deluxetable}{lrrllllllrrp{2cm}p{2cm}}
\tabletypesize{\scriptsize}\tablecaption{Plausible candidate parameters}\tablehead{
    Candidate &  $Kp$ &      $P$ &       $T_0$ & $T_{14}$ &  $(R_P/R_*)^2$ &     $R_*$ &    $M_*$ & \Teff & \logg & In C5? &       Comment &                                        Ephemeris matching \\
    & & [d] & [BJD - 2454833] & [d] & [ppm] & [$R_\odot$] & [$M_\odot$] & [K] & [cgs] \\
    }\startdata
 211397844.01          & 13.292 & 16.168833 & 3274.81174 & 0.1331 & 4046 & $5.653 \pm 0.284$ & $2.750 \pm 0.454$ & $5186 \pm 135$ & $3.371 \pm 0.090$ &  &   &          \\
211492384.01       & 12.375 &  0.635328 & 3262.97422 & 0.0543 & 1315 & $1.531 \pm 0.047$ & $1.286 \pm 0.096$ & $6420 \pm 300$ & $4.176 \pm 0.050$ &  &                                somewhat V-shaped       &                                                          \\
 211503824.01       & 11.246 &  2.047195 & 3263.17910 & 0.0789 & 255 & $0.958 \pm 0.013$ & $1.000 \pm 0.046$ & $6150 \pm 110$ & $4.476 \pm 0.023$ &  &                              slightly asymmetric       &                                                          \\
 211543616.01       & 15.6 &  1.518232 & 3263.37623 & 0.1284 & 356389 & $1.209 \pm 0.074$ & $0.914 \pm 0.073$ & $5510 \pm 210$ & $4.237 \pm 0.062$ &  &                                         V-shaped       &                                                          \\
 211544257.01       & 12.242 &  1.630130 & 3263.45081 & 0.0952 & 46964 & $1.602 \pm 0.047$ & $1.286 \pm 0.082$ & $6190 \pm 230$ & $4.136 \pm 0.041$ &  &                 somewhat V-shaped; spotted star?       &                                                          \\
 211546613.01       & 15.982 &  2.007472 & 3263.89429 & 0.0377 & 16989 & $0.473 \pm 0.012$ & $0.495 \pm 0.016$ & $3790 \pm 70$ & $4.783 \pm 0.013$ &  &                                         V-shaped       &                                                          \\
 211611139.01\tablenotemark{a}       & 12.474 &  1.882213 & 3263.60257 & 0.0402 & 5678 & $1.568 \pm 0.140$ & $1.355 \pm 0.106$ & $6590 \pm 380$ & $4.178 \pm 0.081$ &  &                                somewhat V-shaped       &                                                          \\
\enddata\label{tab:med}
\tablenotetext{a}{Parameters not in EPIC; classified using \texttt{isochrones} as described in Sec.~\ref{sec:stellar}.}
\end{deluxetable}

\clearpage
\end{landscape}

\begin{landscape}
\LongTables 
\begin{deluxetable}{lrrllrlll}
\tabletypesize{\scriptsize}\tablecaption{Systems with Secondary Eclipses}\tablehead{
    Candidate &  $Kp$ &      $P$ &       $T_0$ & $T_{14}$ &  $(R_P/R_*)^2$ &        In C5? &                    Comment &                                         Ephemeris matching \\
    & & [d] & [BJD$_{TDB}$ - 2454833] & [d] & [ppm] & &
}\startdata
211397774.01  &   15.649  &    0.967865  &   3262.83095  &    0.1047  &          149235  &  &             even-odd -- half-period  &                                                     \\
 211402878.01  &   14.498  &   25.236580  &   3277.84155  &    0.1427  &          131088  &  &                                       &                                                     \\
 211408138.01  &   12.740  &   10.337864  &   3267.82296  &    0.2954  &           50598  & yes &                                       &                                                     \\
 211409299.01  &   11.951  &   13.919301  &   3263.16370  &    0.2036  &          126507  & yes &                                       &                                                     \\
 211411891.01  &   12.776  &    0.954863  &   3262.53608  &    0.0928  &          122533  &  &                                       &                                                     \\
 211422822.01  &   15.664  &    0.554633  &   3262.67870  &    0.0547  &           26770  &  &                                       &                                                     \\
 211425822.01  &   14.888  &    2.153178  &   3263.18086  &    0.1114  &          161675  &  &                                       &                                                     \\
 211429934.01  &   14.147  &    0.565934  &   3262.47647  &    0.0817  &           76907  &  &                                       &                                                     \\
\enddata\label{tab:eb}
\end{deluxetable}

\clearpage
\end{landscape}

\LongTables 
\begin{deluxetable}{lrrllp{2cm}}
\tabletypesize{\scriptsize}\tablecaption{Systems showing periodic variability}\tablehead{
    Candidate &  $Kp$ &     $P$ [d] &          In C5? &          Comment &                  Ephemeris matching 
}\startdata
 211404813.01 & 16.261 & 0.560055 &    &                                 & \\
211405917.01 & 16.664 & 0.53736 &  yes  &                                 & \\
211417284.01 & 14.388 & 0.912345 &    &                                 & \\
211419593.01 & 14.084 & 0.544223 &    &                                 & \\
211422471.01 & 9.95 & 1.694853 &    &     & \\
211432103.01 & 10.249 & 0.933482 &  yes  &              likely even-odd   & \\
211432905.01 & 17.747 & 0.565384 &  yes  &                                 & \\
211433054.01 & 13.798 & 0.61983 &  yes  &                                 & \\
\enddata\label{tab:var}
\end{deluxetable}

\end{document}